\documentclass[12pt]{article}
\usepackage[polish]{babel}
\newcommand{\bd}{\begin{displaymath}}
\newcommand{\ed}{\end{displaymath}}
\newcommand{\be}{\begin{equation}}
\newcommand{\ee}{\end{equation}}
\newcommand{\bc}{\begin{center}}
\newcommand{\ec}{\end{center}}
\begin{document}
\begin{titlepage}
\begin{center}
\Large Backreaction of excitations on a domain wall.
\end{center}
\vspace{5 mm}
\bc
by
\ec
\bc
\large Robert Pe"lka
\ec
\vspace{5 mm}
\bc
Jagellonian University, Institute of Physics 
\ec
\bc
Reymonta 4, 30-059 Cracow, Poland
\ec
\vspace{7 mm}
\bc
\large Abstract
\ec
In this paper we investigate backreaction of excitations on a planar domain 
wall in a real scalar field model. The backreaction is investigated in the cases of homogeneous, 
plane wave and wave packet type excitations. We find that the excited domain 
wall radiates. The method of calculating backreaction for the general forms of 
excitations is also presented. 

\end{titlepage}

\section{Introduction}
The presence of topological defects in the field theoretical models with a 
degenerate vacuum is an important aspect of the structure of these models. 
Dynamics of topological defects is to be 
extracted from nonlinear equations describing evolution of fields they are
composed of. The results, eventhough it is a formidable task to get them, 
are of great interest for particle physics (e.g. dynamics of a flux-tube 
in QCD [1]), for field theoretical cosmology (e.g. cosmic strings [2], [3]) and for 
condensed matter physics (e.g. domain walls in magnetics, vortices in 
superconductors or in superliquid helium, defects in liquid crystals [4], 
[5], [6], [7], [8]).

This paper is devoted to dynamics of domain walls governed by a Poincar\`e 
invariant wave equation. Domain walls appear in 
the models with the nontrivial zeroth homotopy group of the vacuum manifold.
In the papers [9], [10], [11] two main approaches to the dynamics of domain walls have been 
presented. The first one is the polynomial approximation, the second one is the 
expansion in the width of the wall. Calculations made in the framework of these 
two methods indicate existence of an oscillating component in
the width of the domain wall. This suggests that the domain wall can radiate.
The radiation was also observed in computer simulations [12].

The problem we adress ourselves to is connected with this special aspect 
of dynamics of domain walls, namely the radiation. We calculate the radiation with the help 
taken from [13], [14] where analogous was considered in the abelian Higgs model. 
The method consists of two steps. 
The first one is to find the excitations of a static domain wall. 
They are investigated in the linear approximation and 
treated as small corrections to the basic domain wall field which is
localized on the wall. The second step consists of looking for the effects of  
this excitations on the evolution of the domain wall, i.e. a backreaction. 
This procedure can be treated as the expansion of the domain wall field 
in the amplitude of the excitation A (A$\ll$1). The zeroth order term is then
the static, exact planar domain wall solution $\phi_0$, the first order term is the excitation 
$\phi_1$ and the second order term is the backreaction
\be
\phi=\phi_0+A\phi_1+A^2\phi_2+O(A^3).
\ee
The terms of the higher order can be interpreted as more complicated effects,
e.g. the third order term as the selfinteraction of the excitation.
We shall discuss three main cases of the excitations, namely the most simple of 
them - the homogenous excitation, the second case - a plane wave type excitation 
and the last one - a localized excitation. 
In this paper we shall consider the dynamics of the domain wall in the simplest 
scalar field model which has a potential with two degenerate minima. 

The plan of our paper is the following. In the next Section we present the 
calculation of the backreaction of the homogeneously excited domain wall.
In Section 3 we describe the method of calculating backreaction for the 
general forms of excitations. Section 4 is devoted to the detailed analysis
of the backreaction in the cases of plane wave and wave packet type excitations.
In Section 5 we summarize the main points of our work.

\section{Radiation from a homogeneously excited domain wall}
We consider the planar domain wall in the model defined by the Lagrangian
density,
\be
\mathcal{L}=-\frac{1}{2}\eta_{\mu\nu}\partial^{\mu}\Phi\partial^{\nu}\Phi-\frac{
\lambda}{2}({\Phi}^2-{\upsilon}^2)^2,
\ee
where $\eta_{\mu\nu}=$diag$(-1,1,1,1)$  and $\lambda , \upsilon $
are positive constants. The corresponding evolution equation for the scalar 
field $\Phi$ has the form:
\be
\partial^{\mu}\partial_{\mu}\Phi-2\lambda({\Phi}^2-{\upsilon}^2)\Phi=0.
\ee
Exact, static solution representing the domain wall localised around the 
 $(x^1 ,x^2)$-plane is given by the formula:
\be
\phi_0 =\upsilon\tanh(\alpha x^3),
\ee
where $\alpha=\sqrt{\lambda{\upsilon}^2}.$

Let us rescale the scalar field $\Phi$ and the space-time coordinates 
$x^{\mu}$,
\be
\begin{array}{c}

\Phi=\upsilon\phi ,\\
\tilde{x}^{\mu}=\alpha x^{\mu} ,
\end{array}
\ee
where $\phi(\tilde{x}^{\mu})$ and $\tilde{x}^{\mu}$ are dimensionless.
Moreover, for $\tilde{x}^0 $ and $\tilde{x}^3 $ we shall use the following 
notation, 
\be
\begin{array}{c}
\tilde{x}^0=\tau, \\
\tilde{x}^3=\xi.
\end{array}
\ee
In the rescaled variables the evolution equation and the static solution 
take the form:
\be
\begin{array}{c}
\tilde{\partial}^{\mu}\tilde{\partial}_{\mu}\phi-2({\phi}^2-1)\phi=0 , \\ \\
\phi_0=\tanh(\xi) ,
\end{array}
\ee
where $\tilde{\partial}^{\mu}=\frac{\partial}{\partial\tilde{x}_{\mu}}$.
The first step of our considerations is to find excitations of the 
planar domain wall. As it was stated in the Introduction the excitations
 are considered as small corrections to the basic static domain wall 
solution. Thus the corresponding equation for the excitation field $\phi_1$ 
we shall obtain as the linear approximation to the initial evolution equation.
Inserting the expansion (1) into Eq.(7) and keeping the terms of the first order
in the expansion parameter A we obtain:
\be
\tilde{\partial}^{\mu}\tilde{\partial}_{\mu}\phi_1 -2(3{\phi_0}^2-1)\phi_1=0.
\ee
The planar domain wall distinguishes the direction perpendicular 
to the wall plane, given in our case by the coordinate lines of $\xi $. Small 
correction $\phi_1 $ do not change this asymmetry. 
Therefore we assume a special form of 
the perturbation in which dependence on the coordinate $\xi$ is separeted,
\be
\phi_1=\phi(\xi)\chi(\tau,\tilde{x}^1 ,\tilde{x}^2).
\ee
Inserting the above Ansatz into Eq.(8) we obtain the system of two equations 
for the functions $\psi$ and $\chi$ with a separation constant c,
\be
\begin{array}{l}
\frac{d^2 \psi}{d\xi^2}-6{\phi_0}^2\psi+2\psi=-c\psi, \\ \\
\tilde{\partial}^a \tilde{\partial}_a \chi=c\chi ,
\end{array}
\ee
where a=0,1,2. From that follows a restiction for constant c. If c
were less than zero the solution of the second equation would increase with
 the time infinitely and so wouldn't meet the main requirement of the  
perturbative calculation. 

Let us solve the first equation 
of the system (10). It can be transformed into the following form:
\be
-\frac{1}{2}\frac{d^2\psi}{d\xi^2}-\frac{3}{\cosh^2\xi}\psi=(\frac{c}{2}-2)\psi.
\ee
It is the special case of Schr\"odinger equation with generalized P\"oschel-Teller 
potential.
Solutions of this equation can be found in [15]. There exist 
two bound states enumerated by n=0,1,
\be
\begin{array}{lll}
n=0 &\psi_0(\xi)=\frac{1}{\cosh^2\xi}, &c_0=0, \\ \\
n=1 &\psi_1(\xi)=\frac{\sinh\xi}{\cosh^2\xi}, & c_1=3.
\end{array}
\ee
Both are localized on the wall in the sense that they exponentially 
decrease for large $\mid\xi\mid$. The first, even solution $\psi_0(\xi)$ can
be interpreted as a small displacement of the domain wall,
\be
\phi=\phi_0+\psi_0\chi=\tanh\xi+\frac{1}{\cosh^2\xi}\chi\simeq\tanh(\xi+\chi).
\ee
This is the zero mode related to the translational symmetry
of the model. 
The excitation given by the second, odd function $\psi_1$ does not posses 
such interpretation. This solution we shall accept as the proper bound state of the domain wall. 

The second equation of the system (10) is of the wave type. In this Section 
we shall consider the special case of excitation, 
homogeneous on the whole wall and given by the solution independent of the 
coordinates $\tilde{x}^1,\tilde{x}^2$,
\be
\chi(\tau)=A\cos(\sqrt{3}\tau+\delta) ,
\ee
where $\delta$ is a constant phase which we shall put equal to zero, and A is a constant
amplitude. As mentioned above the described procedure assumes that A is sufficiently small. 
Finally, the field of the homogeneously excited planar domain 
wall is given by the formula:
\be
\phi=\tanh\xi+A\frac{\sinh\xi}{\cosh^2\xi}\cos(\sqrt{3}\tau).
\ee
From this formula one can see that the excitation introduces periodic changes of 
the thickness of the domain wall.

The second step in our procedure is to find the backreaction. Inserting the 
expansion (1) into the Eq.(7) and keeping all terms of order $A^2$
we obtain the equation for the backreaction $\phi_2$,
\be
\tilde{\partial}^{\mu}\tilde{\partial}_{\mu}\phi_2-2(3{\phi_0}^2-1)\phi_2=
6\phi_2{\phi_1}^2.
\ee
It is an inhomogeneous equation with the r.h.s. including the square of the excitation 
component. We will denote the inhomogeneous term by $N=6\phi_0
{\phi_1}^2$. For a homogeneous excitation we are dealing with,  $N$ is a 
function of the two variables $(\xi,\tau)$. Inserting the formulae for the 
functions $\phi_0$ and $\phi_1$ we obtain:
\be
N=N_1+N_2=3A_2\frac{{\sinh^3}\xi}{{\cosh^5}\xi}+3A^2\frac{{\sinh^3}\xi}
{{\cosh^5}\xi}\cos(2\sqrt{3}\tau).
\ee
We shall solve Eq.(16) in two steps, considering each of the two parts of the
inhomogeneous term $N$ separately. At the first step we consider the equation:
\be
\frac{1}{2}\frac{d^2\phi_2}{d\xi^2}-(3\tanh^2\xi-1)\phi_2=\frac{3}{2}A^2
\frac{\sinh^3\xi}{\cosh^5\xi}.
\ee
The general solution of this equation can be found by the standard Green's 
function technique, see e.g. [16]. As the two linearly independent solutions 
of the homogeneous part of Eq.(16) we take:
\be
\begin{array}{l}
f_1(\xi)=\frac{1}{\cosh^2\xi}, \\ \\
f_2(\xi)=\frac{1}{8}\sinh(2\xi)+\frac{3}{8}\tanh\xi+\frac{3}{8}\frac{\xi}
{\cosh^2\xi}.
\end{array}
\ee
As the Green's function we take
\be
G(\xi,x)=2f_1(x)f_2(\xi)\theta(\xi-x)-2f_1(\xi)f_2(x)[\theta(\xi-x)-\theta(-x)].
\ee
The Green's function was choosen in such a manner as to obey the condition
\be
G(\xi=0,x)=0 \ \ \ \ \bigwedge x\in R.
\ee
Such a choice ensures that an inhomogenuity won't produce any displacement
of the domain wall as a whole. The general solution of Eq.(18) has the form
\be
\phi_2(\xi)=af_1(\xi)+bf_2(\xi)+A^2\int_{-\infty}^{+\infty}G(\xi,x)h(x)dx ,
\ee
where 
\bd
h(x)=\frac{3}{2}\frac{\sinh^3 x}{\cosh^5 x},  
\ed
and $a,b$ are constants.
Formula (22) gives:
\be
\phi_2(\xi)=A^2[c_1(\xi)f_1(\xi)+c_2(\xi)f_2(\xi)],
\ee
\be
\begin{array}{l}
c_1(\xi)=\frac{a}{A^2}-2\int_{0}^{\xi}f_2(x)h(x)dx, \\ 
c_2(\xi)=\frac{b}{A^2}+2\int_{-\infty}^{\xi}f_1(x)h(x)dx.
\end{array}
\ee
The function $f_1$ is even while $f_2$ is odd. Because $f_2(\xi)$ exponentially
grows for $\xi\rightarrow\pm\infty$ the coefficient function $c_2$ has to 
vanish in this limit. Therefore  we have to put $b=0$ while $a$ is still 
arbitrary. But we put $a$ to be zero too because keeping it nonzero would amount 
to including uninteresting solution of the homogeneous equation. 
In the second step we have to solve the Eq.(16) with the second inhomogenous 
term $N_2$ containing a periodic time-dependence. 

The perturbatively obtained Eq.(16) for the beackreaction has radiation type 
solutions which do not vanisch for large $\xi$ as will be shown further. 
In this case we adopt the Helmholtz condition which states that for $\mid\xi
\mid\rightarrow\infty$ only outgoing radiation waves are present.

Hence we consider the following form of the solution:
\bd
\phi_2=\frac{1}{2}A^2[\varphi_{+}(\xi)\exp{(-i2\sqrt{3}\tau)}+\varphi_{-}(\xi)\exp{(i2\sqrt{3}\tau)}], 
\ed
where the positive and negative frequency components are related by complex conjugation,
\bd
\varphi_{-}=[\varphi_{+}]^{*}
\ed
This Ansatz leads to the following equation for the functions $\varphi_{\pm}$,
\be
\left[\frac{1}{2}\frac{d^2}{d\xi^2}+(4+\frac{3}{\cosh^2\xi})\right]\varphi_{\pm}(\xi)=\frac{3}{2}
\frac{\sinh^3\xi}{\cosh^5\xi} ,
\ee
which we solve analogously as Eq.(18). As two linearly independent 
solutions we take 
\be
\begin{array}{l}
g_1(\xi)=\cos(2\sqrt{2}\xi)-\sqrt{2}\tanh\xi\sin(2\sqrt{2}\xi)+\frac{1}{2}
\frac{\cos(2\sqrt{2}\xi)}{\cosh^2\xi}, \\ \\
g_2(\xi)=\sin(2\sqrt{2}\xi)+\sqrt{2}\tanh\xi\cos(2\sqrt{2}\xi)+\frac{1}{2}
\frac{\sin(2\sqrt{2}\xi)}{\cosh^2\xi}.
\end{array}
\ee
The Green's function is given by the formula
\be
G(\xi,x)=\frac{1}{6\sqrt{2}}[g_2(\xi)g_1(x)\theta(\xi-x)-g_1(\xi)g_2(x)(
\theta(\xi-x)-\theta(-x))].
\ee
The general solution has the form 
\be
\varphi_{\pm}(\xi)=[d_1(\xi)+a_{\pm}]g_1(\xi)+[d_2(\xi)+b_{\pm}]g_2(\xi),
\ee
where 
\be
d_1(\xi)=-\frac{1}{6\sqrt{2}}\int^{\xi}_{0}g_2(x)h(x)dx ,
\ee
\be
d_2(\xi)=\frac{1}{6\sqrt{2}}\int^{\xi}_{-\infty}g_1(x)h(x)dx,
\ee
and $a_{\pm},b_{\pm}$ are constants.

In order to satisfy the Helmholtz condition imposed on the solution, we have 
to find solutions with the appriopriate asymptotics given by the formula below
\be
\varphi_{\pm}(\xi\rightarrow\pm\infty)\sim\exp{[\pm i2\sqrt{2}\mid\xi
\mid]}.
\ee
We consider separately the regions $\xi>0$ and $\xi<0$ and choose the 
solutions which will have the proper 
asymptotic behaviour given by the formula (31). Next, we impose the matching 
conditions at the point $\xi=0$, i.e. the continuity conditions of the solution 
and its first derivative. The solutions found in this way are given by the formula
\be
\varphi_{\pm}=d_1(\xi)g_1(\xi)+[d_2(\xi)\pm id_1(\infty)]g_2(\xi) ,
\ee
where 
\bd
d_1(\infty)=\lim_{\xi\rightarrow +\infty}d_1(\xi)
\ed
and is finite.
Finally, the asymptotic form of the solution is the following
\be
\phi_2(\xi\rightarrow\pm\infty)\sim\pm\sqrt{3}d_1(\infty)A^2\cos(2\sqrt{2}\mid\xi\mid-2\sqrt{3}\tau\pm\beta),
\ee
where $\beta=\arctan{\sqrt{2}} , \beta\in(0,\frac{\pi}{2})$.

Let us summarize results of our calculations. The full solution consisting 
of the static domain wall, the excitation and backreaction is the following,
\be
\begin{array}{l}
\phi=\tanh\xi+A\frac{\sinh\xi}{\cosh^2\xi}\cos(\sqrt{3}\tau)+A^2[c_1(\xi)
f_1(\xi)+c_2(\xi)f_2(\xi)]+ \\ \\ +\frac{1}{2}A^2[\varphi_{+}(\xi)\exp{(-i2\sqrt{3}\tau)}+\varphi_{-}(\xi)exp{(i2\sqrt{3}\tau)}].
\end{array}
\ee
Backreaction consists of two terms. The first one, independent of time,  
consists of the static reaction of the domain wall to the homogeneous 
excitation while the second one depends on time. Its asymptotics is given 
by the formula (33) and describes the radiation with the energy fluxes given 
by the Poynting vectors:
\be
\vec{S}_{\pm}=6\sqrt{3}d^2_1(\infty)A^4\upsilon^2\alpha^2\sin^2(k^{\mu}
_{\pm}\tilde{x}_{\mu}\pm\beta)\vec{k}_{\pm},
\ee
where 
\be
k^{\mu}_{\pm}=(2\sqrt{3},0,0,\pm 2\sqrt{2}).
\ee
and the signs $\pm$ correspond to the limits $+\infty$ and $-\infty$ respectively.

\section{The method for the general forms of excitations}
Our goal in this Section is to present the method of calculating backreaction 
for a general form of excitation $\phi_1$ defined by the formula:
\be
\phi_1=\psi(\xi)\chi(\tau,\tilde{x}^1,\tilde{x}^2) ,
\ee
where
\bd
\psi(\xi)=\frac{\sinh\xi}{\cosh^2\xi},
\ed
and the function $\chi$ is any bounded solution of the wave equation
\be
\tilde{\partial}^a\tilde{\partial}_a\chi=3\chi.
\ee
Let us recall the equation for the backreaction,
\be
\tilde{\partial}^{\mu}\tilde{\partial}_{\mu}\phi_2-2(3\tanh^2\xi-1)\phi_2=
6\phi_0{\phi_1}^2.
\ee
The first step is to define the operator $\hat{L}_{\xi}$ by the formula:
\be
\hat{L}_{\xi}=\frac{d^2}{d\xi^2}+\frac{6}{\cosh^2\xi}.
\ee
Then the Eq.(39) takes the form:
\be
\tilde{\partial}^a\tilde{\partial}_a\phi_2+\hat{L}_{\xi}\phi_2-4\phi_2=
6\phi_0{\phi_1}^2 .
\ee
The second step is to solve the eigenvalue problem for this operator. We shall 
find all  solutions of the equation:
\be
\hat{L}_{\xi}\psi_{\lambda}(\xi)=\lambda\psi_{\lambda}(\xi),
\ee
where $\lambda$ is the eigenvalue and $\psi_{\lambda}(\xi)$ - the normalized
 eigenfunction corresponding to this eigenvalue. The 
backreaction $\phi_2$ we treat as the expansion in the eigenfunctions of $\hat{L}_{\xi}$
given by the formula:
\be
\phi_2(\xi,\tilde{x}^a)=\sum\!\!\!\!\!\!\!\!\int_{\lambda}a_{\lambda}(\tilde{x}^a)
\psi_{\lambda}(\xi) ,
\ee
where the coefficients $a_{\lambda}$ depend on the coordinates 
$\tilde{x}^a$. Inserting this expansion into the Eq.(41) we obtain
\be
\sum\!\!\!\!\!\!\!\int_{\lambda'}\psi_{\lambda'}(\xi)[\tilde{\partial}^a\tilde
{\partial}_a+\lambda'-4]a_{\lambda'}(\tilde{x}^a)=N(\xi,\tilde{x}^a) ,
\ee
where $N(\xi,\tilde{x}^a)=6\phi_0{\phi_1}^2$.
Multiplying the equation above by the eigenfunction $\psi_{\lambda}(\xi)$ and 
integrating over the full range of variation of $\xi$ and using the ortonormality 
condition for the system of the eigenfunctions $\psi_{\lambda}(\xi)$ we obtain 
the system of differential equations for the coeffcient functions $a_{\lambda}
(\tilde{x}^a)$,
\be
(\tilde{\partial}^a\tilde{\partial}_a+\lambda-4)a_{\lambda}(\tilde{x}^a)=
h_{\lambda}(\tilde{x}^a) ,
\ee
where
\bd
h_{\lambda}(\tilde{x}^a)=\int_{-\infty}^{+\infty}d\xi\psi_{\lambda}(\xi)
N(\xi,\tilde{x}^a).
\ed
The last step in this procedure is to solve Eq.(45). It may be done by the 
standard method of the Green's function. Let us denote by $G_{\lambda}(\tilde{
x}^a)$ the retarded Green's function of the operator, 
\be
\hat{D}_{\lambda}=\tilde{\partial}^a\tilde{\partial}_a+\lambda-4.
\ee
Then the solution of the inhomogeneous equations (45) is given by the formula
\be
a_{\lambda}(\tilde{x}^a)=\int G_{\lambda}(\tilde{x}^a-\tilde{x'}^a)h_{\lambda}(
\tilde{x'}^a)d\tilde{x'}^a.
\ee
In this solution we have dropped the homogeneous part
 because it does not contain any information about the excitation.

The results of the calculations are the following. 
The spectrum of the operator $\hat{L}_{\xi}$ consists of two parts. The discrete
part, for which $\lambda>0$, contains two eigenfunctions, 
\be
\begin{array}{ll}
\psi_1=\frac{\sqrt{3}}{2}\frac{1}{\cosh^2\xi}, & \lambda_1=4, \\
\psi_2=\sqrt{\frac{3}{2}}\frac{\sinh\xi}{\cosh^2\xi}, & \lambda_2=1.
\end{array}
\ee
The continuous part, for which $\lambda\leq0$, includes twice degenerate
subspaces corresponding to the eigenvalues  $\lambda=-k^2 (k\in R_{+})$ 
spanned by the eigenfunctions: 
\be
\begin{array}{l}
{\psi_k}^{(1)}(\xi)=n(k)[(k^2-2)\cos(k\xi)-3k\sin(k\xi)\tanh\xi+\frac{3
\cos(k\xi)}{\cosh^2\xi}] , \\
{\psi_k}^{(2)}(\xi)=n(k)[(k^2-2)\sin(k\xi)+3k\cos(k\xi)\tanh\xi+\frac{3\sin(
k\xi)}{\cosh^2\xi}] ,
\end{array}
\ee
where
\bd
n(k)={[\pi k (k^2+1)(k^2+4)]}^{-\frac{1}{2}}.
\ed
Therefore we have to find the retarded Green's function for the following operators:
\be
\begin{array}{ll}
\hat{D}_1=\tilde{\partial}^a\tilde{\partial}_a & (\lambda=4), \\
\hat{D}_2=\tilde{\partial}^a\tilde{\partial}_a-3 & (\lambda=1), \\
\hat{D}_k=\tilde{\partial}^a\tilde{\partial}_a-k^2-4 & (\lambda=-k^2).
\end{array}
\ee
They are obtained by the Fourier transform method and given by the formulae:
\be
\begin{array}{l}
G_1(\tilde{z}^a)=-\frac{1}{2\pi}\frac{\theta{\tilde{z}^0}}{\sqrt{{(\tilde{z}^0)
}^2-{(\tilde{z})}^2}}, \\     \\
G_2(\tilde{z}^a)=-\frac{\theta({\tilde{z}}^0)}{2\pi}\int_{0}^{\infty}dK K 
\frac{\sin(\sqrt{K^2+3}\tilde{z}^0)}{\sqrt{K^2+3}}J_0 (K\tilde{z}), \\    \\
G_k(\tilde{z}^a)=-\frac{\theta({\tilde{z}}^0)}{2\pi}\int_{0}^{\infty}dK K
\frac{\sin(\sqrt{K^2+k^2+4}\tilde{z}^0)}{\sqrt{K^2+k^2+4}}J_0(K\tilde{z}) ,
\end{array}
\ee
where 
\bd
\begin{array}{l}
\tilde{z}^a=\tilde{x}^a-\tilde{x}'^a, \\
\tilde{z}=\sqrt{{(\tilde{z}^1)}^2+{(\tilde{z}^2)}^2} .
\end{array}
\ed
The procedure presented above enables us to find the backreaction for the general form of 
excitation of domain wall. We shall use it in the next Section.

\section{The backreaction for the plane wave and wave packet excitations}

In this section we analyse the backreaction of the excitation of the
plane wave and wave packet type. In the first case $\chi$ is given by formula:
\be
\chi=A\cos(\omega(k_0^1,k_0^2)\tau-k_0^1\tilde{x}^1-k_0^2\tilde{x}^2),
\ee
where $\omega(k_0^1,k_0^2)=\sqrt{(k_0^1)^2+(k_0^2)^2+3}$.

In the second case we consider the approximate solution of Eq.(38):
\be
\chi\cong A\exp[-\frac{{(\tilde{x}^1)}^2+{(\tilde{x}^2)}^2}{{\Lambda}^2}]
\sin(\sqrt{3}\tau)=Aw(\tilde{x})\sin(\sqrt{3}\tau) ,
\ee
where A and $\Lambda$ are constants. We assume that $\Lambda\gg 1$ and that
 the wave packet has momentum cutoff $k\sim{\Lambda}^{-1}$. Then we may neglect
for the finite time interval $0\leq\tau\leq\sim\Lambda$ the spreading out of 
the wave packet, which is of course present in exact wave packet solutions of
Eq(38).

\underline{1.The plane wave case.}

\noindent The inhomogeneous terms in the backreaction equation has the following form: 
\be 
N^{(1)}(\tau,\tilde{x}^1,\tilde{x}^2,\xi)=3 A^2\frac{{\sinh}^3\xi}{{\cosh}^5\xi}[
1-\cos(2\omega(k_0^1,k_0^2)\tau-2k_0^1\tilde{x}^1-2k_0^2\tilde{x}^2)]
\ee
It is convenient to pass to the Fourier transform with respect to the coordinates 
$\tilde{x}^1 ,\tilde{x}^2 ,\tau$:
\be
\begin{array}{c}
\phi^{(1)}_{2}(\xi,\tilde{x}^1,\tilde{x}^2,\tau)= \\   \\
\frac{1}{{(2\pi)}^{\frac{3}{2}}}\int\!\!dk^1\int\!\!dk^2\int\!\!d\omega
\exp{[-i\omega\tau+ik^1\tilde{x}^1+ik^2\tilde{x}^2]}\hat{\phi}^{(1)}_{2}
(\xi,k^1,k^2,\omega) ,
\end{array}
\ee
\be
\begin{array}{c}
N^{(1)}(\xi,\tilde{x}^1,\tilde{x}^2,\tau)=  \\    \\  \frac{1}{{(2\pi)}^{\frac{3}{2}}}\int
\!\!dk^1\int\!\!dk^2\int\!\!d\omega \exp{[-i\omega\tau+ik^1\tilde{x}^1+ik^2\tilde{x}^2]}
\hat{N}^{(1)}(\xi,k^1,k^2,\omega).
\end{array}
\ee
In this case,
\be 
\begin{array}{c}
\hat{N}^{(1)}(\xi,k^1,k^2,\omega)=3{(2\pi)}^{\frac{3}{2}}A^2\frac{{\sinh}^3\xi}
{{\cosh}^5\xi}[\delta (\omega)\delta(k^1)\delta(k^2)\\ \\-\frac{1}{2}\delta(\omega
+2\omega(k^1_0,k^2_0))\delta(k^1+2k^1_0)\delta(k^2+2k^2_0) \\  \\ 
-\frac{1}{2}\delta(\omega-2\omega(k^1_0,k^2_0))\delta(k^1-2k^1_0)
\delta(k^2-2k^2_0)].
\end{array}
\ee
The equation for the Fourier transform $\hat{\phi}^{(1)}_{2}$ has the following
form:
\be
[\omega^2-k^2+\frac{d^2}{d\xi^2}-2(3\tanh^2\xi-1)]\hat{\phi}^{(1)}_{2}(\xi,
k^1,k^2,\omega)=\hat{N}^{(1)}(\xi,k^1,k^2,\omega).
\ee
It is clear from the above equation that the solution has the form:
\be
\begin{array}{c}
\hat{\phi}^{(1)}_{2}(\xi,k^1,k^2,\omega)=A^2{(2\pi)}^{\frac{3}{2}}
[{\varphi}^{(1)}_{0}(\xi)\delta(\omega)\delta(k^1)\delta(k^2)\\  \\ 
   -\frac{1}{2}
{\varphi}^{(1)}_{-}(\xi)\delta(\omega+2\omega({k}^{1}_{0},{k}^{2}_{0}))
\delta(k^1+2{k}^{1}_{0})\delta(k^2+2{k}^{2}_{0}) \\ \\ 
-\frac{1}{2}{\varphi}^{(1)}_{+}(\xi)\delta(\omega-2\omega(k^1_0,k^2_0))
\delta(k^1-2k^2_0)\delta(k^2-2k^2_0)].
\end{array}
\ee
The negative and positive frequency components are related by complex 
conjugation,
\be
\varphi^{(1)}_{-}=[\varphi^{(1)}_{+}]^{*} 
\ee
while $\varphi^{(1)}_{0}$ is real valued.
The functions $\varphi^{(1)}_{0} ,\varphi^{(1)}_{\pm}$ have to satisfy
the following equations:
\be
\left[\frac{1}{2}\frac{d^2}{d\xi^2}+\frac{3}{\cosh^2\xi}-2\right
]\varphi^{(1)}_{0}=\frac{3}
{2}\frac{\sinh^3\xi}{\cosh^5\xi} ,
\ee
\be
\left[\frac{1}{2}\frac{d^2}{d\xi^2}+\frac{3}{\cosh^2\xi}+4\right]\varphi^{(1)}_{\pm}=
\frac{3}{2}\frac{\sinh^3\xi}{\cosh^5\xi}.
\ee
The function $\varphi^{(1)}_{0} $corresponding to the frequency $\omega=0$
contains the information about the static backreaction of the domain wall 
while two remaining functions describe the dynamic backreaction. We shall concentrate 
on these two functions only. They must satisfy identical conditions as the functions 
$\varphi_{\pm}$ in Section 2 and they also satisfy the same equation, see formulae (25 - 31). 
Therefore, the solutions are given by the formula (32). 
The asymptotic form of the radiation part of the backreaction is then the following:
\be
\begin{array}{c}
\phi^{(1)}_2(\tilde{x}^1,\tilde{x}^2,\xi,\tau)\sim \\    \\ \mp\sqrt{3}A^2d_1(\infty)
\cos{[2k^1_0\tilde{x}^1+2k^2_0\tilde{x}^2+2\sqrt{2}\mid\xi\mid-
2\omega(k^1_0,k^2_0)\tau\pm\beta]} ,
\end{array}
\ee
where $\beta=\arctan{\sqrt{2}} , \beta\in(0,\frac{\pi}{2})$.
The corresponding wave vectors have the components: 
\be
(k^{\mu}_{\pm})=(2\omega(k^1_0,k^2_0),2k^1_0,2k^2_0,\pm 2\sqrt{2}) ,
\ee
where the signs $\pm$ correspond to the limit in the $\pm\infty$ respectively.
The energy fluxes due to these waves are given by the Poynting vectors: 
\be
\vec{S}_{\pm}=6d_1^2(\infty)A^4\upsilon^2\alpha^2\omega(k^1_0,k^2_0)\sin^2(k^{\mu}_{\pm}\tilde{x}_
{\mu}\pm\beta)\vec{k}_{\pm}.
\ee

\underline{2.The wave packet case.}

\noindent Calculations of backreaction are carried out in the analogous steps 
as in plane wave case. Instead of formulae (54),(57) we have now: 
\be
N^{(2)}(\xi,\tilde{x}^1,\tilde{x}^2,\tau)=3A^2\frac{\sinh^3\xi}{\cosh^5\xi}w^2(
\tilde{x})[1-\cos(2\sqrt{3}\tau)] ,
\ee
\be
\hat{N}^{(2)}(\xi,k^1,k^2,\omega)=3A^2\frac{\sinh^3\xi}{\cosh^5\xi}w(k)[\delta(\omega)
-\frac{1}{2}\delta(\omega+2\sqrt{3})-\frac{1}{2}\delta(\omega-2\sqrt{3})] ,
\ee
where
\bd
w(k)\equiv\frac{1}{2\pi}\int\!\!d\tilde{x}^1\int\!\!d\tilde{x}^2
\exp{[-ik^1\tilde{x}^1-ik^2\tilde{x}^2]}w^2(\tilde{x}) ,
\ed
\bd
k=\sqrt{(k^1)^2+(k^2)^2}.
\ed
We next obtain the same equation for the Fourier transform $\hat{\phi}^{(2)}_2(\xi
,k^1,k^2,\omega)$ of $\phi^{(2)}_2(\xi,\tilde{x}^1,\tilde{x}^2,\tau)$:
\be
\left[\omega^2-k^2+\frac{d^2}{d\xi^2}-2(3\tanh^2\xi-1)\right]\hat{\phi}^{(2)}_2=
\hat{N}^{(2)}.
\ee
Analogously to the previous case the solution can be written as
\be
\hat{\phi}^{(2)}_2=A^2\frac{\sinh^3\xi}{\cosh^5\xi}w(k)[\varphi^{(2)}_0(\xi)
\delta(\omega)-\frac{1}{2}\varphi^{(2)}_{-}(\xi)\delta(\omega+2\sqrt{3})-
\frac{1}{2}\varphi^{(2)}_{+}(\xi)\delta(\omega-2\sqrt{3})].
\ee
In the exact solution the functions $\varphi^{(2)}_0 ,\varphi^{(2)}_{\pm}$ 
dependent on $k$, but 
in the case at hand the expression in square brackets on the left hand 
side of Eq.(68) can be simplified. Namely, we may neglect the term $k^2$ because 
of the cutoff $\Lambda\gg 1$. This simplification 
leads to the same set of equations (61),(62) for the functions $\varphi^{(2)}_0 , 
\varphi^{(2)}_{\pm}$ as in the previous case and the solution given by the 
formula (69) is then the approximate one. The asymptotic conditions (31) remain 
unchanged. Thus the asymptotic form of the radiation part of the backreaction 
is the following:
\be
\phi^{(2)}_2(\xi\rightarrow\pm\infty,\tilde{x}^1,\tilde{x}^2,\tau)\sim
\mp\sqrt{3}d_1(\infty)A^2\cos(2\sqrt{2}\mid\xi\mid-2\sqrt{3}\tau\pm\beta)w^2(\tilde{x}).
\ee
The corresponding wave vector has the components:
\be
k^{\mu}_{\pm}=(2\sqrt{3},0,0,\pm 2\sqrt{2}).
\ee
The energy flux is given by the formula:
\be
\vec{S}_{\pm}=6\sqrt{3}d^2_1(\infty)A^4\upsilon^2\alpha^2w^4(\tilde{x}
)\sin^2(k^{\mu}
_{\pm}\tilde{x}_{\mu}\pm\beta)\vec{k}_{\pm}.
\ee
\section{Remarks}
1.Let us summarize the main points of our work. We have presented
the calculations of the backreaction in the cases of the homogeneous, plane 
wave and wave packet type excitations of the domain wall. We also have described the 
method enabling us to analyse the more general cases of the excitations.
The main result of our work is the existence of the long range component 
in the backrection which is interpreted as the radiation from the excited 
domain wall. The frequency of the radiation, given by $\phi_2$,(cf. formulae (63),
(70)), is twice of that of the excitation function $\chi$, (cf. formulae (52),(53)).

\noindent
2.The idea of the expansion in the amplitude of excitation, which all the 
calculations were based on, was applied to the simplest model of the real 
scalar field and the potential $V(\phi)=\frac{\lambda}{2}(\phi^2-\upsilon^2)^2$.
It seems that without much trouble this method could be applied also to the other field-theoretical 
models containing domain wall configuration. On the other 
hand, we should remember that this method is based on the linear 
approximation, what implies that it is reliable for small amplitudes of the 
excitations only. In order to consider stronger excitations we have to 
work out another method which will take into account the nonlinearity of the evolution 
equation in a better manner. One could for instance use the polynomial approximation 
in the vicinity of the domain wall and the proper asymptotics at the 
infinity and smoothly match them in the intermediate region.

\large Acknowledgements

\noindent
\normalsize
I would like to thank prof. H.Arod"z for his help, interest and stimulating 
discussions. I am also grateful to dr L.Hadasz for his help in the editing 
of this paper.

\Large References

\normalsize
[1] See, e.g., M.Baker, J.S.Ball, F.Zachariasen, Phys.Rep. \underline{209}, 73 (1991).

[2] T.W.B.Kibble, J.Phys. \underline{A9}, 1387 (1976).

[3] A.L.Vilenkin, Physics Reports \underline{121}, 263 (1985).

[4] See, e.g. , J.Slonczewski, in Physics of Defects (Les Houches Session 

\hspace{4 mm} XXXV, 1980). North-Holland Publ. Comp., Amsterdam, 1981.

[5] R.P.Huebener, Magnetic Flux Sructures in Superconductors.

\hspace{4 mm} Springer-Verlag, Berlin - Heidelberg -New York, 1979.

[6] R.J.Donnelly, Quantized Vortices in HeliumII. Cambridge University 

\hspace{4 mm} Press, Cambridge, 1991.

[7] S.Chandrasekhar and G.S.Ranganath, Adv. Phy. \underline{35}, 507 (1986).

[8] W.H."Rurek, Phys.Rep. \underline{276}, 177 (1996).

[9] H.Arod"z, Phys.Rev. \underline{D52}, 1082 (1995).
 
[10] H.Arod"z, A.L.Larsen, Phys.Rev. \underline{D49}, 4154 (1994).

[11] H.Arod"z, Nucl.Phys. \underline{B450}, 174 (1995).

[12] L.M.Widrow, Phys.Rev. \underline{D40}, 1002 (1989).

[13] H.Arod"z, L.Hadasz, Phys.Rev. \underline{D54}, 4004 (1996).

[14] H.Arod"z, L.Hadasz, Phys.Rev.  \underline{D55}, 942 (1997).

[15] S.Fl\"ugge, ''Practical quantum mechanics I'' (Springer-Verlag, 1971).

[16] G.Korn,T.Korn, ''Sprawocznik po matematike'' (Moskwa, 1970).

\end{document}